\documentclass[aps,prl,reprint,groupedaddress]{revtex4-1}
\usepackage{graphicx}
\usepackage{amssymb,amsfonts,amsmath,footmisc}
\usepackage{natbib}
\begin{document}
\title{Critical Thickness Ratio for Buckled and Wrinkled Fruits and Vegetables}

\author{Hui-Hui Dai}
\author {Yang Liu}
\affiliation{Department of Mathematics, City University of Hong Kong,
83 Tat Chee Avenue, Hong Kong, PR China}

\date{\today}

\begin{abstract}
Fruits and vegetables are usually composed of exocarp and sarcocarp and they take a variety of shapes when they are ripe.
Buckled and wrinkled fruits and vegetables are often observed. This work aims at establishing the geometrical constraint
for buckled and wrinkled shapes based on a mechanical model. The mismatch of expansion rate between the exocarp and sarcocarp can produce
a compressive stress on the exocarp. We model a fruit/vegetable with exocarp and sarcocarp as a hyperelastic layer-substrate structure
subjected to uniaxial compression. The derived bifurcation condition contains both geometrical and material constants. However, a careful
analysis on this condition leads to the finding of a critical thickness ratio which separates the buckling and wrinkling modes, and
remarkably, which is independent of the material stiffnesses. More specifically, it is found that if the thickness ratio is smaller than this
critical value a fruit/vegetable should be in a buckling mode (under a sufficient stress); if a fruit/vegetable in a wrinkled shape the thickness ratio
is always larger than this critical value. To verify the theoretical prediction, we consider four types of buckled
fruits/vegetables and four types of wrinkled fruits/vegetables with three samples in each type. The geometrical parameters for the 24 samples are measured
and it is found that indeed all the data fall into the theoretically predicted buckling or wrinkling domains. Some practical applications based on this
critical thickness ratio are briefly discussed.
\end{abstract}


\maketitle

\section{Background and model}
Many vegetables and fruits contain exocarp and sarcocarp and usually exocarp is stiffer than sarcocarp in order to protect it.
There are many different morphologies for fruits and vegetables, and in particular wrinkled and globally buckled shapes are often observed, e.g. wrinkled pumpkins
and buckled cucumbers (see Figs. 6 and 7).  Why a fruit/vegetable takes the final shape when they ripe may be due to many different factors
during the growth process. But, mechanical forces alone can play a very important role for determining the geometry of vegetables and fruits
(cf. \cite{ds}). Now, it has been understood that the out layer of plant meristems often expands faster than the inner one, leading to
to the whole structure under compression on the interface (see \cite{ds} and \cite{jh}). In fact, a newborn fruit or
vegetable has a smooth surface, and only after the certain period of growth, the morphology features occur and remain since (cf. \cite{jmps}
to see details). Over the past decades or so, many authors have used purely mechanical models to study patterns in biological objects, e.g., fingerprint formation
by using the Von Karman's equations (\cite{kn2004}, \cite{kn2005}) and pattern formation in plants through shell instability (\cite{jam}) and shapes of sympetalous flowers through growth of a thin elastic sheet (\cite{bmmm}).

Actually, it has been known for some time instabilities can lead to a variety of patterns in a mechanical system. For example, for a thin layer coated to a compliant substrate and a core/shell system, various highly
ordered patterns can occur due to the mismatched deformation(\cite{nature}, \cite{science}, \cite{hhs}, \cite{prl}, \cite{prl1}). Also for a thin shell or plate, wrinkling will occur when the structure is under indentation(\cite{prl2}, \cite{prl3}). Since ordered patterns also appear in many planets, those works certainly
give the motivations to use suitable mechanical structures which resemble the plants to capture those patterns in order to gain certain understanding of their formation. In \cite{prl4}, the authors consider the stretching of a thin sheet, when the in plane strain $\gamma$ reaches a critical value, wrinkling will occur at the longitudinal direction due to compression. Their model can be applied to estimate the elastic properties of our skin or some fruit qualitatively.
In \cite{pnas} and \cite{jmps}, the authors approximated some fruits and vegetables as  spheroidal and ellipsoidal core/shell systems. Under suitable stresses, they analyzed the
post-bifurcation states to understand the different morphologies as the four dimensionless parameters (the shape factor, the thickness ratio of core/shell,
the modulus ratio of shell/core and the growth stress ratio) vary. The numerical simulations successfully captured the similar morphologies of different fruits and vegetables by specifying
the geometrical and material parameters in certain ranges. The work by Yin {\it et al.}(\cite{pnas} and \cite{jmps}) made it clear that geometrical parameters have a great influence on the number of wrinkles in wrinkled fruits/vegetables. Here, we examine the influence of the geometrical parameters on the final shapes of fruits and vegetables with exocarp and sarcocarp from a different aspect.  The aim is to show that there exists a critical thickness ratio of sarcocarp and exocarp, which separates a buckled shape and a wrinkled shape, independent of the stiffnesses of sarcocarp and exocarp. For a fruit/vegetable with exocarp and sarcocarp, we model it as a structure of a layer bonded to a substrate. As discussed before, the stress can play a vital role in the morphology of a fruit/vegetable. As an idealization, we suppose that the structure is subjected to a uniform compressive stress everywhere, which is equivalent to applying a uniaxial compression at both ends. The mechanical model is depicted in Fig. 1, where $\bar{b}$, $\bar{a}$ and $\bar{l}$ denote the thicknesses of the layer and substrate and the length respectively. We point out that in this paper we only consider the onset to a buckled or wrinkled shape, not the morphology beyond. Of course, the surface of a fruit or vegetable is not planar, but as pointed out in \cite{liu}, it resembles the layer-substrate structure and it is expected that the theory can be applied with reasonable accuracy.

Both exocarp and sarcocarp are modelled as Saint-Venant materials, for which the strain energy takes the form
\begin{equation}
\Phi=\frac{\lambda}{2}(\textrm{Tr}\textbf{E})^2+\mu
(\textrm{Tr}\textbf{E}^2),
\end{equation}
where
$\textbf{E}=\frac{\textbf{F}^T\textbf{F}-\textbf{I}}{2}$
is the Green strain tensor ($\textbf{F}$ is the deformation gradient), and $\lambda$ and $\mu$
are the Lam\'{e} constants. It is supposed that the Lam\'{e} constants for the layer and substrate are different. For convenience, a bar on a quantity is referred to one for the layer and a quantity without a bar is referred to that for the substrate. For example, $\bar{\lambda}, \bar{\mu}$ represent the Lam\'{e} constants of the layer. For further simplicity, we assume that both layer and substrate have the same Poisson's ratio and their Young moduli $\bar{E}$ and $E$ are different.

The onset to a buckled shape or a wrinkled one can be determined by a linear bifurcation analysis of the current model. For a single hyperelastic layer/plate under compression, the linear bifurcation analysis was given in \cite{dr1994}. For a layer bonded to a half-space, we refer to \cite{bigoni}, \cite{caifu1999} and \cite{caifu2000} for the corresponding analysis. Here, in our model the substrate is of finite thickness and we present the linear bifurcation analysis in the appendix. According to such an analysis, the bifurcation condition can be written in the following form:
\begin{equation}
\textbf{f}
(r,Y,\nu,\frac{\bar{b}}{\bar{l}},m_1,n)=0,
\end{equation}
where the function $\textbf{f}$ is in the form of a determinant given in the supporting information, $Y=\frac{\bar{E}}{E}$ is the ratio of Young moduli, $\nu$ is the Poisson's ratio, $\frac{\bar{b}}{\bar{l}}$ is the aspect ratio of the layer, $r=\frac{\bar{a}}{\bar{b}}$ is the
thickness ratio of substrate and layer, $n$ is the wave number and $m_1$ is the critical stretch. Once the geometrical parameters $\frac{\bar{b}}{\bar{l}}$, $r$ and material parameters $Y$ and $\nu$ are specified, for a given wave number this equation determines the critical stretch (the critical stress can then be easily determined).  For this layer-substrate structure, the first mode (for which the critical stretch is the largest, i.e. the critical compressive stress is smallest) can be either a buckling mode or a wrinkling mode. For the buckling modes, usually the critical stretch decreases as the wave number increases so the first mode is for $n=1$. However, for $n=1$ the buckled shape is not symmetric but for buckled fruits/vegetables the shape is roughly symmetrical about the middle. Based on such a consideration, we only study $n\ge 2$ modes.

In Fig. 2, we fix the values of $Y$, $\nu$ and $\frac{\bar{b}}{\bar{l}}$ and plot the critical stretch curves versus the thickness ratio for $n=2, 3, 11$ and $12$.  We point out that the curves for all other modes (which are not shown) are always below either $n=2$ curve or $n=11$ curve and thus the first mode can only be the $n=2$ mode or $n=11$ mode, depending on whether the thickness ratio is smaller or larger than the value $r_c$ shown in the figure. The bifurcation analysis in the supporting information also gives the eigenfunction for each mode, and based on which we can plot the eigen shape of the structure for a given mode.  For the parameters chosen in Fig. 2, $r_c=18.0565$.  Then, for $r=18$ the first mode is $n=2$ mode and $r=18.1$ the first mode is $n=11$ mode. The eigen shapes of the structure for these two values of thickness ratio are shown in Fig. 3. One can observe that for $r<r_c$ the structure is in a buckling mode while for $r>r_c$ it is a wrinkling mode. Thus, the critical thickness ratio $r_c$ separates buckling and wrinkling modes. To further analyze $r_c$, we observe that $r_c$ corresponds to the intersection point of the $n=2$ curve and an $n\ge 3$ curve which has the largest stretch value (in Fig. 2 it happens that this curve is $n=11$). Therefore, $r_c$ can be determined from the following three equations:
\begin{flalign}
&\nonumber\textbf{f}(r,Y,\nu,\frac{\bar{b}}{\bar{l}},m_1,n=2)=0,\\\nonumber
&\textbf{f}(r,Y,\nu,\frac{\bar{b}}{\bar{l}},m_1,n)=0 (n\geq 3),\\
&\frac{\partial \textbf{f}}{\partial n}(r,Y,\nu,\frac{\bar{b}}{\bar{l}},m_1,n)=0(n\geq 3).\label{three}
\end{flalign}
Once $\frac{\bar{b}}{\bar{l}}$, $Y$ and  $\nu$ are given, $r_c$, the wave
number $n$ and the critical stretch $m_1$ can be found. It turns out that the value of Poisson's ratio has little influence on $r_c$ and thus from now on we fix its value to be $\nu=0.1$. Then, for a given $\frac{\bar{b}}{\bar{l}}$, the above three equations yield a relation between $r_c$ and the ratio of Young moduli. For $\frac{\bar{b}}{\bar{l}}=0.01$, we plot the $r_c - Y$ curve in Fig. 4. For a given $Y$, if the thickness ratio is above or below this curve, the structure is in a wrinkling mode or buckling mode. Another intrinsic feature is that this curve has a global minimum at  $r_c=r_m$. Thus, under the condition that there is a sufficient compressive stress, if $r<r_m$ the structure is always in a buckling mode and if the structure is a wrinkling mode one must have $r>r_m$. Another importance of the existence of $r_m$ is that it is independent of the ratio of Young moduli. Actually, for $r_m$ we should have $\frac{\partial r_c}{\partial Y}=0$, which leads to
\begin{flalign}
\nonumber &\frac{\partial\mathbf{f}}{\partial Y}(r_m,Y,\nu,\frac{\bar{b}}{\bar{l}},m_1,n=2)\\
&+\frac{\partial\mathbf{f}}{\partial m_1}(r_m,Y,\nu,\frac{\bar{b}}{\bar{l}},m_1,n=2)\frac{\partial m_1}{\partial Y}=0,\label{newone}
\end{flalign}
where
\begin{flalign}
\frac{\partial m_1}{\partial Y}=
-\frac{\frac{\partial^2 \mathbf{f} }{\partial n \partial Y}\frac{\partial \mathbf{f}}{\partial n}
+\frac{\partial^2 \mathbf{f}}{\partial n^2 }\frac{\partial \mathbf{f}}{\partial Y}}
{\frac{\partial^2 \mathbf{f}}{\partial n^2 }\frac{\partial \mathbf{f}}{\partial m_1}+
\frac{\partial^2 \mathbf{f} }{\partial n \partial m_1}\frac{\partial \mathbf{f}}{\partial n}}.
\end{flalign}
This equation together with the previous three equations in $(3)$ determine a relation between $r_m$ and
$\frac{\bar{b}}{\bar{l}}$. We point out that this relation is purely geometrical and is independent of the material parameters of the layer and substrate. Probably, the existence of this critical ratio $r_m$ and which is only related to the aspect ratio of the layer is the major finding of the present theoretical analysis.

We plot the $r_m$ curve as $\frac{\bar{b}}{\bar{l}}$ varies in Fig. 5. This
$r_m$ curve divides the whole $\frac{\bar{b}}{\bar{l}}-r$ plane into two
parts. Based on the discussions  below $(3)$ and this figure, we can conclude that, no matter what the material parameters are, for a given aspect ratio of the layer, if the thickness ratio is below this $r_m$ curve the structure is in a buckling mode (with a sufficient compressive stress) and if the structure is in a wrinkling mode the thickness ratio must be above this $r_m$ curve.

\section{Measured data and discussions}

The theoretical analysis on a layer-substrate structure demonstrates the criticalness of the thickness ratio $r_m$ in determining the buckled or wrinkled shape. Now, we shall examine this ratio for a number of fruits and vegetables to see whether it can provide the correct prediction. The moduli of fruits and vegetables are difficult to measure. However, both the thickness ratio of exocarp and sarcocarp and the aspect ratio of exocarp are geometrical parameters, which can be measured without much difficulty. More specifically, we choose eight types of vegetables and fruits (each with three samples) for measurement. The wrinkled fruits/vegetables include large pumpkin, small pumpkin, chayote and Korean melon (see Fig. 6) and buckled fruits/vegetables include banana, emperor banana, zucchini and cucumber (see Fig. 7). We mention that all samples are ripe fruits/vegetables. More properly, one should use samples just before they start buckling or wrinkling but it is a difficult task to acquire them. So here, the assumption is that the thickness ratio and aspect ratio of ripe fruits/vegetables are not much different from those near buckled or wrinkled shapes at the growing process.

Since our model is two-dimensional, we shall take certain cross section of a fruit/vegetable for measurement. For those buckled fruits/vegetables, we take a longitudinal cross-section and then cut the half-part
averagely along the latitudinal direction (see Fig. 7). Then, this half cross section can be considered as a layer-substrate structure. It is pretty straightforward to measure the thicknesses of the exocarp and sarcocarp and the length (although for the thin exocarp the measurement may have some error and we assume that our measurement has an error of $10\%$ for the thickness of exocarp). The measured data are given in Table 1. For those wrinkled fruits/vegetables, we take a latitudinal cross section and also cut it into half (see Fig. 6). It is not difficult to measure the thicknesses of exocarp and sarcocarp.  But, since the cross section is not flat, the corresponding length of the structure in the mechanical model is not obvious. Here, as a simplification we take it as the average of the outer and inner boundary lengths. The data for these wrinkled fruits/vegetables are given in Table 2.

We plot the measured data in the $\frac{\bar{b}}{\bar{l}} - r_m$ plane together with the critical $r_m$ curve. The results for the buckled fruits/vegetables are shown in Fig. 8. Due to the possible $10\%$ measurement error for the thickness of exocarp, for each sample the data is represented as a line segment. For example, for the three banana samples there are three line segments.  For the four types of buckled fruits/vegetables, there are in total twelve line segments, which, as can be seen, are all below the critical $r_m$ curve. Thus, indeed, the thickness ratio of sarcocarp and exocarp and aspect ratio of exocarp is located in the buckling mode as predicted by the theoretical analysis. In particular, for the samples of banana, emperor banana and cucumber those nine line segments are well below the $r_m$ curve, which implies that even with a bigger measurement error the theory provides the correct prediction. The data for four types of wrinkled fruits/vegetables are shown in Fig. 9. As one can see, all the twelve line segments are located well above the critical $r_m$ curve. In this case, it implies that even the measurement error is bigger than $10\%$ the data still fall into the theoretically predicted wrinkling mode. In summary, the existence of the critical thickness ratio of sarcocarp and exocarp, which separates a buckled shape and a wrinkled shape, is supported by the 24 samples of buckled and wrinkled fruits/vegetables.

\section{Conclusion}
Motivated by understanding the formation of a buckled or wrinkled shape of fruits/vegetables, we
consider a simple mechanical model with a layer-substrate structure under uniaxial compression. A linear bifurcation analysis
leads to the bifurcation condition for the onset to a buckling or wrinkling mode. Some further analysis
on this condition yields a critical thickness ratio (as a function of the aspect ration of the layer).
It seems to be rather remarkable, this critical ratio, which is purely geometrical and independent
of the material parameters, can provide a sufficient condition for a buckled shape (under sufficient compressive stress)
and a necessary condition for a wrinkled shape. That is, for a given aspect ratio of the layer, under sufficient compressive
stress the structure is in a buckled shape if the thickness ratio is smaller than this critical value; if the structure is in a wrinkled
shape the thickness ratio must be larger than this critical value. We measure four types of buckled fruits/vegetables and four types of
wrinkled fruits/vegetables (each with three samples) and the data for all the 24 samples support the theoretical prediction. The study appears
to reveal such a critical thickness ratio is intrinsic for buckled and wrinkled fruits/vegetables.

It should be pointed that the growth of fruits and vegetables is really
complicated, combining together mechanical, biological and biochemical processes(\cite{bn}, \cite{science2003}, \cite{sms}). Nevertheless, a purely mechanical
model may still provide useful insights, as demonstrated here that the samples indeed support the existence of a critical thickness ratio for buckled
and wrinkled fruits/vegetables. At least, the work once again
shows the importance of geometrical parameters in determining the shapes of fruits/vegetables.  As a side product, the results could help in choosing
a fruit/vegeable in our daily life. For example, if a pumpkin has no wrinkles, the thickness ratio of sarcocarp and exocarp should be relatively small (less than $r_m$), which
could imply a relatively thick exocarp or thin sarcocarp. On the other hand, if a pumpkin has wrinkles, the thickness ratio should be larger than $r_m$, which could imply
a relatively thick sarcocarp or thin exocarp. Actually, a further analysis of equation $(2)$ shows that in the wrinkling mode the mode number $n$ (i.e., wrinkle number) is
a decreasing function of the thickness of the layer. Thus, choosing a pumpkin with more wrinkles is more desirable as its exocarp is relatively thin or its
sarcocarp is relatively thick.

\begin{acknowledgments}
The work described in this paper was supported by a GRF grant from Research Grants Council of Hong Kong, HKSAR, China (Project No.: CityU 100911).
\end{acknowledgments}

\appendix
\section{Linear bifurcation analysis}
The bifurcation condition for the mechanical model used in this paper (see Fig. 1) can be obtained by an incremental theory in a similar way as in \cite{dr1994} for
a hyperelastic plate/layer.

The initially stress-free state of the layer-substrate structure
is denoted by $B_0$ and the critical state immediately before buckling/wrinkling is denoted by  $B$. The deformation from $B_0\longrightarrow B$
should be homogeneous, as it is a state caused by a uniform compression. Denoting the deformation
gradient of the substrate arising from $B_0\longrightarrow B$  by $\textbf{F}$, we have
\begin{align}
\textbf{F}=\left(%
\begin{array}{cc}
   m_1 & 0 \\
  0 & m_2 \\
\end{array}%
\right),
\end{align}
where $m_1$ and $m_2$ are the stretches along the $X$-axis and $Y$-axis respectively. Using the traction-free boundary conditions at the bottom, for a Saint-Venant material it is easy to deduce that
\begin{flalign}
m_2=\frac{\sqrt{2\lambda+4\mu-\lambda
m^2_1}}{\sqrt{\lambda+4\mu}}.
\end{flalign}
A similar relation can be obtained for the layer.

Now we superimpose a small deformation with displacement component $(u_1, u_2)$ on $B$, denote
the new state by $B_t$ and also take $B$ as the new reference configuration.  The linearized incremental nominal stress is (see \cite{dr1994})
\begin{flalign}
\tilde{\textbf{S}}_0=A_0\tilde{\textbf{F}},
\end{flalign}
where $\tilde{\textbf{F}}$ is the deformation gradient from $B\longrightarrow B_t$ and
$A_0$, a fourth-order tensor, is the first-order instantaneous elastic moduli referred to $B$. The linearized incremental
governing equations (neglecting the body force) for a static problem, can be written as
\begin{flalign}
\textrm{div} \tilde{\textbf{S}}_0=0,
\end{flalign}
where $\textrm{div}$ is the divergence operator in $B$. Substituting $(8)$ into $(9)$, we have the governing equations in component form
\begin{flalign}
A_{0jilk}u_{k,lj}=0 \quad i,j \in {1,2},
\end{flalign}
where $()_{,i}$ denotes $\frac{\partial}{\partial x_i}()$ and $(x_1, x_2)$ are the rectangular coordinates in $B$.  Note that
the summation convention is adopted unless otherwise stated. $A_{0jilk}$ can be expressed in terms $m_1$ and $m_2$, whose formulas can be found in \cite{ogden1984}. The incremental traction-free boundary condition at the bottom of the substrate gives
\begin{flalign}
A_{021lk}u_{k,l}=0, \quad A_{022lk}u_{k,l}=0, \quad \textrm{at} \quad y=-m_2 \bar{a}.
\end{flalign}

Similar governing equations for the layer and traction-free boundary condition at the top can be also obtained, and here we omit the details.

Also,  it is assumed that the interface between the layer
and substrate is perfectly bonded. As a result, the traction and
displacement should be continuous at the interface, and we have
\begin{flalign}
A_{021lk}u_{k,l}=\bar{A}_{021lk}\bar{u}_{k,l}, \quad
A_{022lk}u_{k,l}=\bar{A}_{022lk}\bar{u}_{k,l}, \quad \textrm{at} \quad y=0,
\end{flalign}
\begin{flalign}
u_1=\bar{u}_1,\quad  u_2=\bar{u}_2, \quad \textrm{at} \quad y=0,
\end{flalign}
where a bar refers to a quantity in the layer. For the two ends, it is assumed that they keep flat during the deformation
and there is no shear stress, which lead to
\begin{flalign}
u_{1y}=0,\quad A_{011lk}u_{k,l}=0, \quad \textrm{at} \quad x=0,\quad m_1\bar{l}.
\end{flalign}

We seek a solution of $(10)$ of the form
\begin{flalign}
u_1=u(y)\textrm{sin}(kx),\quad u_2=v(y)\textrm{cos}(kx).
\end{flalign}
Substituting the above expressions into (10) yields a system of two linear second-order ordinary differential equations for $u(y)$ and $v(y)$. The characteristic equation of the system  has four different complex roots, which are denoted
by $\lambda_1+i\lambda_2$, $\lambda_1-i\lambda_2$,
$-\lambda_1+i\lambda_2$, $-\lambda_1-i\lambda_2$. Then, it is easy to get the following solution:
\begin{flalign}
\nonumber v&=C_1\textrm{sinh}(\lambda_1y)\textrm{cos}(\lambda_2y)
+C_2\textrm{cosh}(\lambda_1y)\textrm{sin}(\lambda_2y)
\\&+C_3\textrm{cosh}(\lambda_1y)\textrm{cos}(\lambda_2y)
+C_4\textrm{sinh}(\lambda_1y)\textrm{sin}(\lambda_2y),
\end{flalign}
\begin{flalign}
\nonumber
u&=C_1(\xi_1\textrm{cosh}(\lambda_1y)\textrm{cos}(\lambda_2y)-\xi_2\textrm{sinh}(\lambda_1y)\textrm{sin}(\lambda_2y))
\nonumber\\&+
C_2(\xi_2\textrm{cosh}(\lambda_1y)\textrm{cos}(\lambda_2y)+\xi_1\textrm{sinh}(\lambda_1y)\textrm{sin}(\lambda_2y))
\nonumber \\&\nonumber+
C_3(\xi_1\textrm{sinh}(\lambda_1y)\textrm{cos}(\lambda_2y)-\xi_2\textrm{cosh}(\lambda_1y)\textrm{sin}(\lambda_2y))\\&+
C_4(\xi_2\textrm{sinh}(\lambda_1y)\textrm{cos}(\lambda_2y)+\xi_1\textrm{cosh}(\lambda_1y)\textrm{sin}(\lambda_2y)),
\end{flalign}
where $\lambda_1$, $\lambda_2$, $\xi_1$ and $\xi_2$ are related to
the material parameters in the Saint-Venant strain energy function, $m_1$ and $k$, and $C_i
(i=1,2,3,4)$ are arbitrary constants.

Similarly, we can also get the general solution of the layer with four constants $(C_5,C_6,C_7,C_8)$ and
the long expressions are omitted.

Substituting $(15)-(17)$ into the conditions $(14)$, we find that $k=\frac{n\pi}{l} (n=1,2,\ldots)$, where $l = m_1 \bar{l}$ and $n$ is the
so-called wave number.

By the use of the four continuity conditions
at the interface (cf. $(12)$, $(13)$), we can represent $(C_5,C_6,C_7,C_8)$
in terms of $(C_1,C_2,C_3,C_4)$, and finally the four traction-free
boundary conditions at the top and bottom yield the following linear algebraic system:
\begin{align}
\textbf{M}\cdot \textbf{C}=0,
\end{align}
where $\mathbf{C}=(C_1,C_2,C_3,C_4)^{\textbf{T}}$ and $\textbf{M}$ is a $4\times4$ matrix and we omit the lengthy expressions for the elements.
In order to get non-trivial solutions for $C_i$
(i=1,$\cdot\cdot\cdot$,4), the determinant of the coefficient
matrix $\textbf{M}$ must vanish, that is,
$\textrm{det}(\textbf{M})=0$, which is the bifurcation condition. For simplicity, we assume that
the layer and substrate have the same Poisson's ratio. Then this determinant is related to the
ratio of Young moduli $Y=\frac{\bar{E}}{E}$, the Poisson's ratio, the thicknesses ratio of the substrate and
layer $r=\frac{\bar{a}}{\bar{b}}$, the stretch $m_1$ and wave number $n$. We denote the bifurcation condition as
\begin{equation}
\textrm{det(\textbf{M})}=:\textbf{f}
(r,Y,\nu,\frac{\bar{b}}{\bar{l}},m_1,n)=0.
\end{equation}
In \cite{liudai},  we consider a similar structure in which the substrate is composed of a Blatz-Ko material.
In particular, we give a complete classification of the parameter domains for different modes and provide a full
asymptotic analysis.

\begin{figure*}\centering
\includegraphics[scale=0.5]{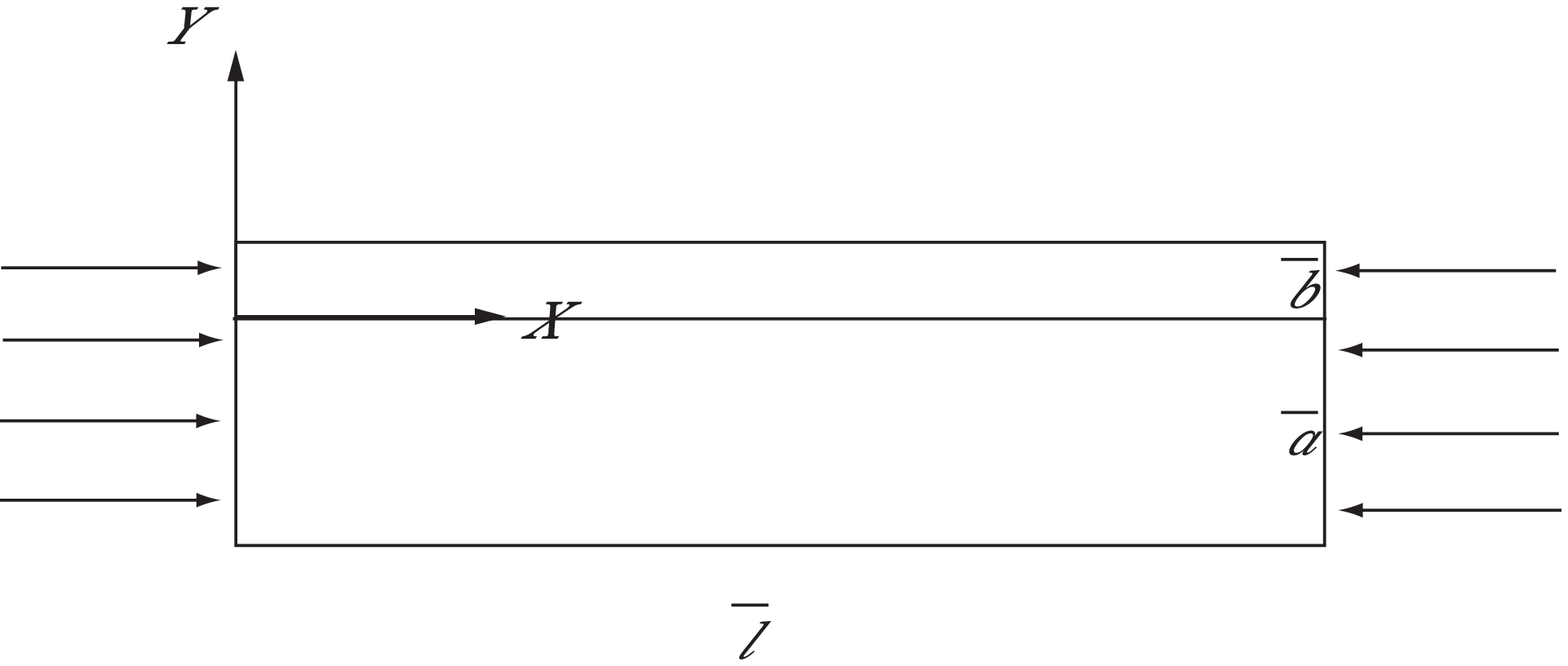}\caption{The geometry of the
model.}\label{fig:geometry}
\end{figure*}

\begin{figure*}\centering
\includegraphics[scale=0.5]{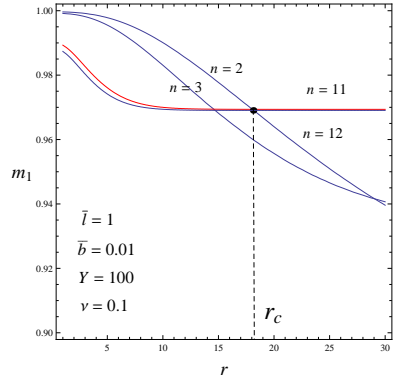}\caption{The curves of the
eigen-stretch as a function of $r$.}\label{fig:eigenstretch}
\end{figure*}

\begin{figure*}
\centering
\includegraphics[scale=0.7]{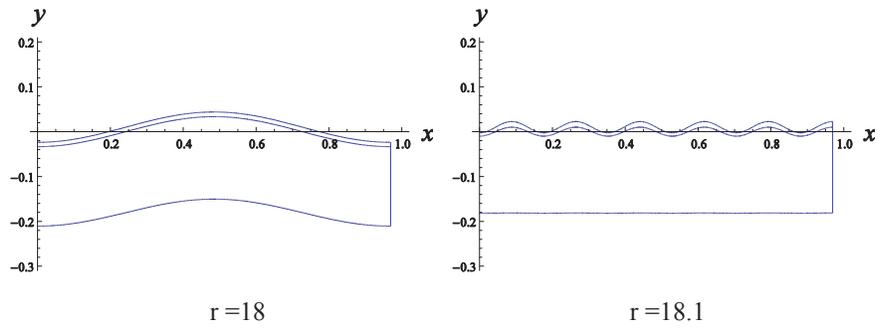}
\caption{Eigen shapes of the structure for two different thickness ratios.
Parameters: $\bar{b}=0.01, Y=100$, $\nu=0.1$ and $\bar{l}=1$.}\label{fig:eigenmode}
\end{figure*}

\begin{figure*}\centering
\includegraphics[scale=0.8]{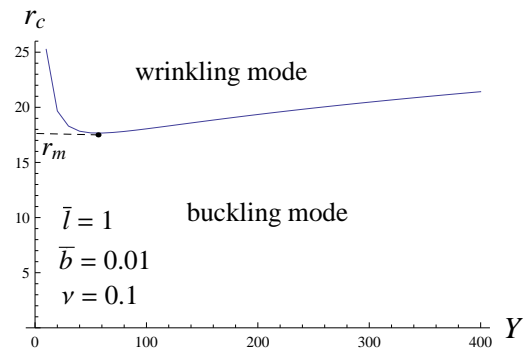}
\caption{The curve of the
critical ratio $r_c$ as a function of $Y$.}\label{fig:rc}
\end{figure*}

\begin{figure*}
\centering
\includegraphics[scale=0.8]{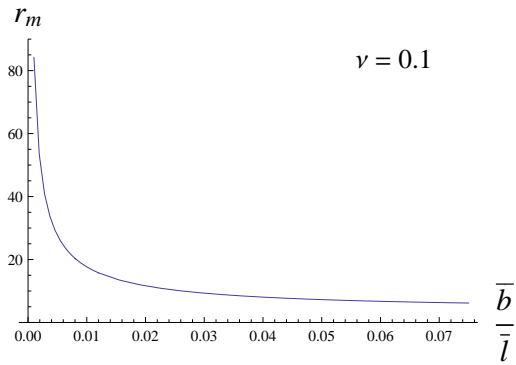}
\caption{The curve of the
critical ratio $r_m$ as a function of the aspect ratio of the layer.}\label{fig:rm}
\end{figure*}

\begin{figure*}\centering
\includegraphics[scale=0.8]{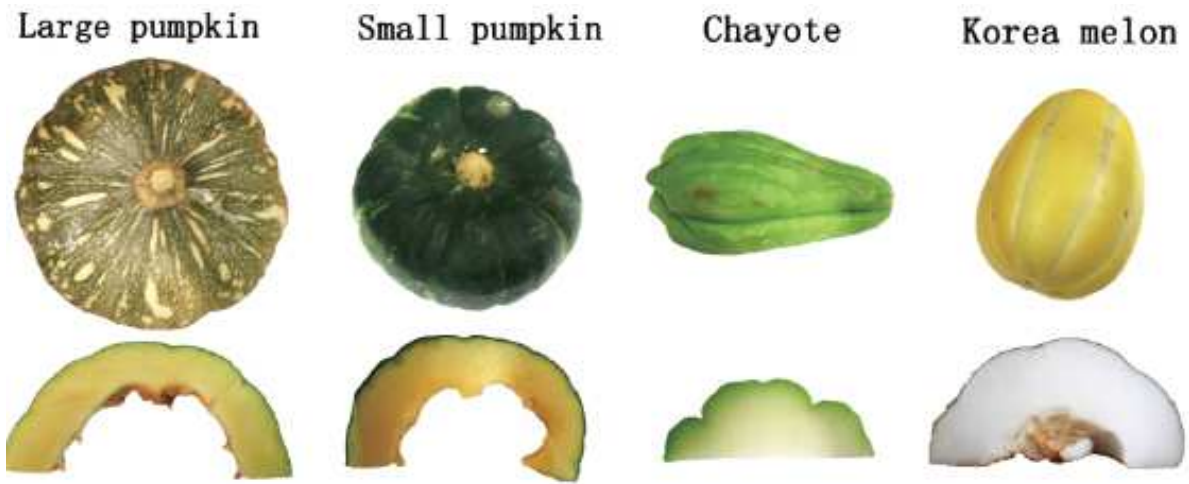}\caption{Measured wrinkled fruits and vegetables.}\label{fig:wrinkling}
\end{figure*}

\begin{figure*}\centering
\includegraphics[scale=0.7]{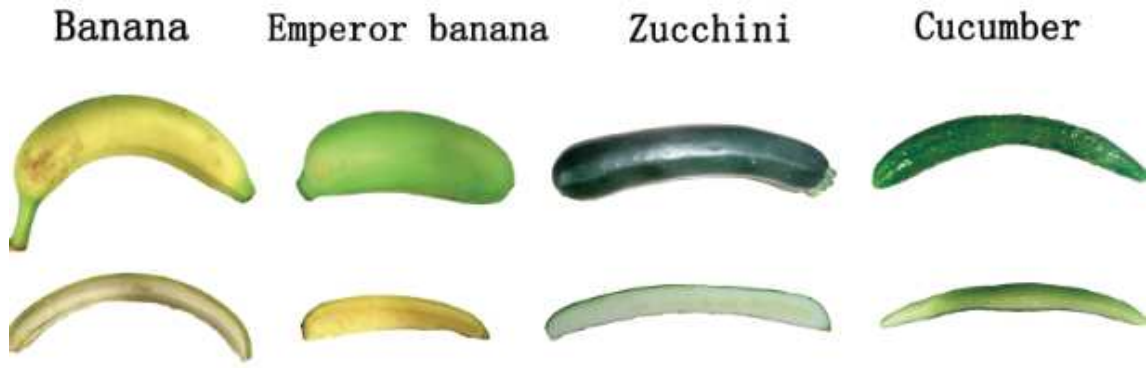}\caption{Measured buckled fruits and vegetables.}\label{fig:buckling}
\end{figure*}

\begin{figure*}\centering
\includegraphics[scale=0.9]{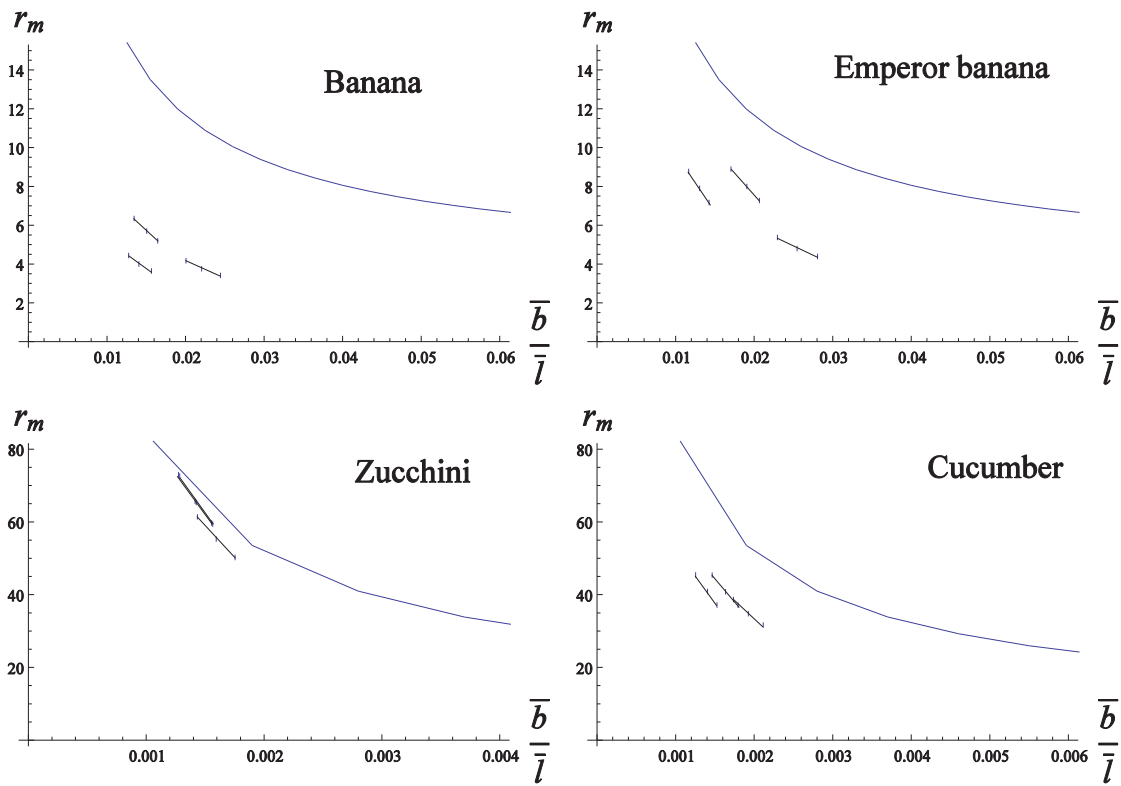}\caption{Measured data for buckled samples and
$r_m$ curve.}\label{fig:buckle}
\end{figure*}

\begin{figure*}\centering
\includegraphics[scale=0.9]{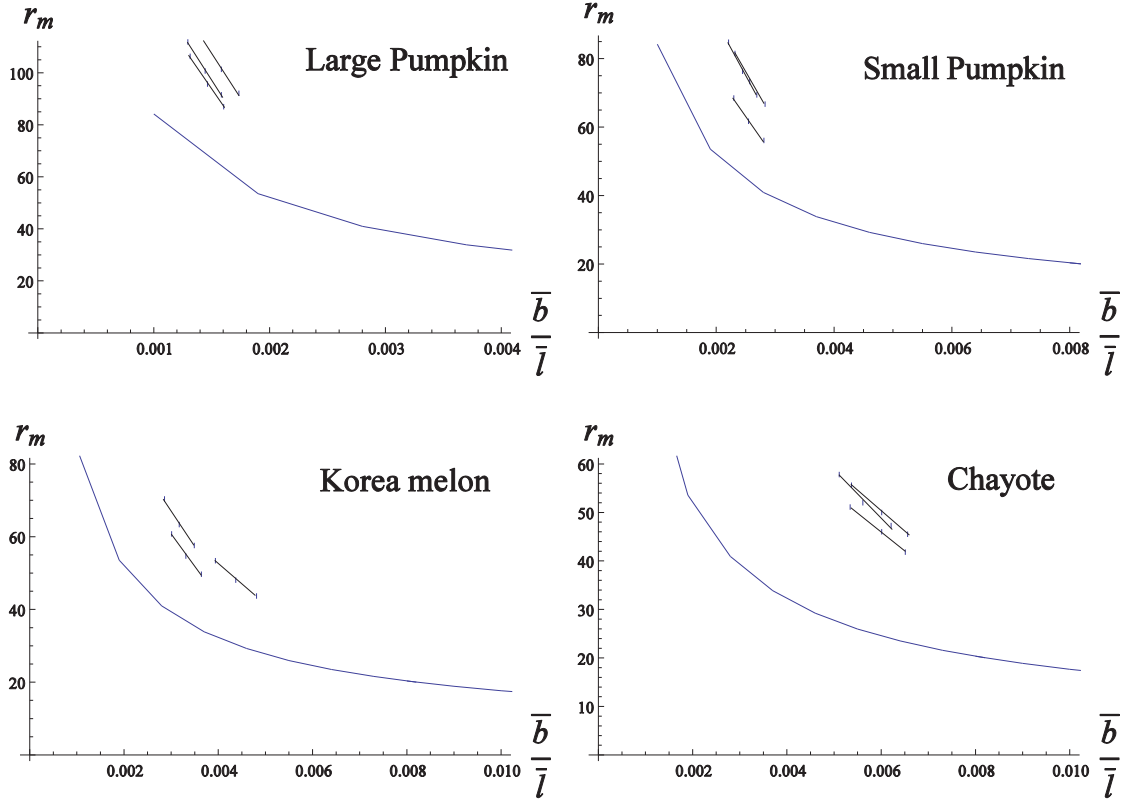}\caption{Measured data for wrinkled samples and
$r_m$ curve.}\label{fig:wrinkle}
\end{figure*}

\begin{table*}[H]
\centering
\caption{The geometrical data of buckled vegetables and fruits}
\begin{tabular*}{\hsize}{@{\extracolsep{\fill}}lrcc}
Name&Thickness of exocarp(cm)&Thickness of sarcocarp(cm) &Length(cm)\cr
\hline
Banana1&0.40&1.50&18.0\cr
\hline
Banana2&0.30&1.20&21.2\cr
\hline
Banana3&0.32&1.83&21.5\cr
\hline
Emperor banana1&0.150&1.20&8.0\cr
\hline
Emperor banana2&0.178&0.86&7.0\cr
\hline
Emperor banana3&0.110&0.87&8.5\cr
\hline
Zucchini1&0.040&2.21&25.1\cr
\hline
Zucchini2&0.030&1.96&21.1\cr
\hline
Zucchini3&0.034&2.23&24.1\cr
\hline
Cucumber1&0.046&1.60&24.0\cr
\hline
Cucumber2&0.048&1.96&29.5\cr
\hline
Cucumber3&0.046&1.88&33.2\cr
\hline
\end{tabular*}
\end{table*}

\begin{table*}[H]
\centering
\caption{The geometrical data of wrinkled vegetables and fruits}
\begin{tabular*}{\hsize}{@{\extracolsep{\fill}}lrcc}
Name&Thickness of exocarp(cm)&Thickness of sarcocarp&Length\cr
\hline
Large Pumpkin1&0.03&3.02&20.9\cr
\hline
Large Pumpkin2&0.03&2.87&20.6\cr
\hline
Large Pumpkin3&0.03&3.04&19.0\cr
\hline
Small Pumpkin1&0.03&2.29&12.3\cr
\hline
Small Pumpkin2&0.03&1.85&11.8\cr
\hline
Small Pumpkin3&0.03&2.20&11.7\cr
\hline
Korea melon1&0.030&1.90&9.48\cr
\hline
Korea melon2&0.036&1.73&8.25\cr
\hline
Korea melon3&0.030&1.64&9.10\cr
\hline
Chayote1&0.05&2.30&8.45\cr
\hline
Chayote2&0.05&2.50&8.40\cr
\hline
Chayote3&0.05&2.60&8.90\cr
\hline
\end{tabular*}
\end{table*}

\end{document}